\documentclass[a4paper]{jpconf}
\usepackage{graphicx}

\newcommand{\beq}{\begin{equation}}
\newcommand{\eeq}{\end{equation}}
\newcommand{\beqa}{\begin{eqnarray}}
\newcommand{\eeqa}{\end{eqnarray}}

\begin{document}
\def\ii{\'\i}

\title{Why pseudo-complex General Relativity? and Applications}

\author{Peter O. Hess}

\address{Instituto de Ciencias Nucleares, Universidad Nacional 
Aut\'onoma de M\'exico, Circuito Exterior, C.U., 
A.P. 70-543, 04510 M\'exico D.F., Mexico \\
and \\
Frankfurt Institute for Advanced Studies, Johann Wolfgang 
Goethe Universit\"at, Ruth-Moufang-Str. 1, 60438 Frankfurt am Main, Germany}

\ead{hess@nucleares.unam.mx}

\begin{abstract}
A brief discussion on the {\it pseudo-complex General Relativity}
is presented. It is shown that this theory is a viable extension
of GR, with deviations centered near to the event horizon.
The theory introduces a dark energy accumulation, due to
the coupling to the central mass.
Predictions of this theory are resumed, as for example the 
structure in an accretion disk, with a dark ring followed by
a bright ring further in. The current {\it Event Horizon
Telescope} observation of M87 is not able to discriminate
between GR and pcGR, due to a low resolution. Further predictions
are also discussed, as the physics of neutron stars,
the redshift at the surface of the star and 
{\it Quasi Periodic Object}.
\end{abstract}

\section{Introduction}

{\it General Relativity} (GR) is one of the best tested
theories in existence. Solar system experiments \cite{will}
confirm the theory, as recent observations of gravitational
waves \cite{gw} and the observation of a black hole shadow
by the {\it Event Horizon Telescope} (EHT) \cite{EHT}.

Nevertheless, one has to keep testing a theory at 
its extreme limits, as for very strong gravitational fields.
Though, the EHT observation is consistent with GR, its
resolution is not sufficient to discriminate theories which
only deviate from GR near the event horizon. Further,
there are conceptual problems in the case of a black hole,
as the singularity in its center, the information loss and 
the existence of an event horizon which separates the inner 
region from the outer one. The existence
of an event horizon is a matter of opinion. 
The event horizon is just a coordinate singularity
and there exist others, well established ones. However, this
one is the result of a strong gravitational fields and it
may bother that the interior of a black hole 
"in the corner of a room" cannot be accessed.

One proposal to extend GR is the {\it pseudo-complex General
Relativity} (pcGR) which makes definite
predictions for the region near the event horizon 
\cite{hess2009,book,universe, PPNP} and also avoids
the event horizon. The theory has no singularity and 
information can get out of the black hole, because of that
rather called a {\it black star}, though, in many respects
this dark star behaves like a black hole. Deviations from
GR are only noticeable near the event horizon. Less massive
object, as neutron stars, behave in pcGR as they do in GR.

In this contribution, I present the motivation for the pcGR 
and why it is a viable proposal for an extension of GR.
Also, some of the predictions made by this theory
will be resumed.

\section{An algebraic extension of General Relativity}

Earlier attempts on extending GR algebraically have been
proposed by A. Einstein \cite{einstein1,einstein2} and
M. Born \cite{born1,born2}. 
Einstein extended the metric to a complex one, where the
real part is the metric used in GR and the imaginary
component is associated to the electromagnetic
energy-momentum tensor. The motivation was to
unify GR with Electrodynamics, but unfortunately
in the limit of small curvature the theory of 
Electrodynamics was not recovered and this approach 
was abandoned. This extension is equivalent
to introducing complex coordinates instead of $x^\mu$
\cite{PPNP}. M. Born's motivation was different: He noted that
coordinates and momenta are not treated equivalently, as
is done in Quantum Mechanics. He introduced the concept of 
{\it complementary} and added a metric term quadratic in the momenta.
In order to observe units, the momentum term carries
a dependence on a {\it minimal length scale factor}. This
renders integrals over momenta finite, another reason why
M. Born advocated this extension. The appearance of a minimal
length scale, as a parameter, has also the 
advantage that Lorentz symmetry is not broken. Later,
E.R. Caianiello \cite{caianiello} modified M. Born's 
length element and the effects of the minimal length were
discussed in \cite{feoli}. 

All these extensions can be summarized under {\it algebraic
extensions}. In an algebraic extension the coordinates
are redefined as

\beqa
x^\mu & \rightarrow & X^\mu ~=~ x^\mu + \sum_ka_k y^\mu_k
\nonumber \\
a_k a_p & = & \sum_q C_{kpq} a_q
~~~. 
\label{eq1}
\eeqa
The algebraic relation in the 
second line in (\ref{eq1}) is the reason to call it{
an {\it algebraic extension}. The meaning of $y^\mu_k$
has to be deduced for each type of extension. 

The simplest example is the {\it complex extension} 
$X^\mu = x^\mu + i y^\mu$, with $i^2=-1$, as proposed in
\cite{mantz1}. Another, not so well known extension, is
to {\it pseudo-complex coordinates} (pc) (also called hyperbolic
coordinates) $X^\mu = x^\mu + I y^\mu$, with $I^2=+1$. 

There are many more possible coordinate systems, 
but the important point is that 
in \cite{kelly} all possible algebraic extensions were
investigated if they contain ghost and/or tachyon solutions,
considered as rendering a theory inconsistent. They
found that only two algebraic versions of the coordinates
are allowed, namely real coordinates, leading to GR,
and pseudo-complex coordinates, leading to a theory called
the {\it pseudo-complex General Relativity}. The main path
in \cite{kelly} is to consider the weak field limit and 
determine the propagators of gravitational waves. In this approach and restricting here to the complex or pseudo-complex
extension only, two types of propagators appear, one with a 
factor of 1 and another with $i^2$ or $I^2$, for the complex
and pseudo-complex case respectively. In the complex case
$i^2=-1$ and refers thus to a ghost solution. Only in the
pseudo-complex case the factor $I^2$ is still 1, rendering
this extension as the only viable one.

For this reason, the pcGR was considered as a viable 
extension to GR. In pcGR the infinitesimal length
square element has the same form as in GR, but now
in terms of the pc-variables:

\beqa
d\omega^2 & = & g_{\mu\nu}(X) dX^\mu dX^\nu
\nonumber \\
g_{\mu\nu} & = & g^R_{\mu\nu} + I g^I_{\mu\nu}
~,~ X^\mu ~=~ x^\mu + I y^\mu
~~~.
\label{eq2}
\eeqa
with, this the length element acquires the form

\beqa
d\omega^2 & = & 
\left\{g^R_{\mu\nu}\left[ dx^\mu dx^\nu +dy^\mu dy^\nu\right]
+ g^I_{\mu\nu}\left[ dx^\mu dy^\nu + dy^\mu dx^\nu\right]\right\}
\nonumber \\
&& 
+ I \left\{ g^I_{\mu\nu} \left[ dx^\mu dx^\nu + dy^\mu dy^\nu
\right] +
g^R_{\mu\nu} \left[ dx^\mu dy^\nu + dy^\mu dx^\nu \right]
\right\}
~~~.
\label{eq3}
\eeqa

Concerning the real part, all formerly mentioned extensions,
mentioned above, can
be accommodated, i.e., it also contains a minimal length.
Because a particle can move only along a real path,
the pseudo-imaginary component of the length element has 
to vanish, which leads to the {\it constraint}

\beqa
g^I_{\mu\nu} \left[ dx^\mu dx^\nu + dy^\mu dy^\nu
\right] +
g^R_{\mu\nu} \left[ dx^\mu dy^\nu + dy^\mu dx^\nu \right]
& = & 0
~~~.
\label{eq4}
\eeqa

To solve (\ref{eq4}) is particular easy in a flat space, with
$g^R_{\mu\nu}=\eta_{\mu\nu}$ and $g^I_{\mu\nu}=0$. The 
constraint then reduces to (skipping a factor of 2)

\beqa
\eta_{\mu\nu} dx^\mu dy^\mu & = & 0
~~~,
\label{eq5}
\eeqa
which is nothing but the standard dispersion relation with
the solution
$y^\mu$ $\sim$ $u^\mu$, i.e., $y^\mu$ is proportional to 
the 4-velocity. For dimensional reasons, a minimal
length scale factor has to be introduced, and using $c=1$
the $y^\mu$ acquires the form

\beqa
y^\mu & = & l u^\mu
~~~.
\label{eq6}
\eeqa
Thus, the $y^\mu$ components are related to the minimal 
length scale.
In this case the length element of E.R. Caianiello can be 
recovered. A more general, approximate 
solution of (\ref{eq4}) is given in \cite{PPNP}.

In order to obtain the extended Einstein equations, one
uses the standard form of the action

\beqa
S=\int dX^4 \sqrt{-g}\left({\cal R}+2\alpha \right)
~~~,
\label{eq6a}
\eeqa
where ${\cal R}$ is the pc-Riemann scalar and $\alpha$
is related to the dark, energy as in GR. For cosmological
models, the $\alpha$ has to be a constant due to translational invariance. However, in a central problem the $\alpha$ may depend
on $r$ and with rotation included also on the
azimuthal angle $\theta$}. In contrast to
GR, all elements in (\ref{eq6a}) are pseudo-complex. For
a mathematical introduction on the analysis of pc-variables
and for further references, please consult \cite{book}.

The action principle $\delta S$ = 0 is applied. In former
publication a modified variational principle was proposed. But as
shown already on the last pages of \cite{book}, including the
constraint (\ref{eq4}) it is equivalent to use the standard
variational principle. {\it Neglecting 
the minimal length} finally
leads to the Einstein equations

\beqa
{\cal R}_{\mu\nu} - \frac{1}{2}g_{\mu\nu}{\cal R}
& = &
8\pi T_{\mu\nu}^\Lambda
~~~.
\label{eq6b}
\eeqa
For more details, please consult \cite{PPNP}. In a current
investigation \cite{l-influence}, the influence of a minimal 
length within pcGR is investigated, with the result that 
only for masses of the order of $l$ appreciable changes 
occur, thus justifying to neglect $l$ for macroscopic 
black holes.
The energy-momentum tensor depends on a couple of parameters 
which leaves space for further assumptions. This leads to
the introduction of dark energy, with the density
$\varrho_\Lambda$.

In addition to the pseudo-complex structure, pcGR assumes
that dark energy accumulates near a black hole
(or a mass in general), but only noticeable near the event
horizon. This is a result of semi-classical calculations
in Quantum Mechanics \cite{birrell,visser}, i.e., dark energy 
is created in a curved space-time back ground. The pcGR
assumes therefore that there is a coupling between the 
central mass and the amount of dark energy. The ansatz
for the dark energy is

\beqa
\varrho_\Lambda & \sim & \frac{B_n}{r^{n+2}}
~~~,
\label{eq7}
\eeqa
with $B_n$ and $n$ being parameters of the theory.

This assumption implies an important principle, namely
that {\it a central mass not only curves space.time but also
changes the vacuum properties of space-time near it}.
This coupling to the dark energy is a consequence of
quantum effects and the study of pcGR therefore can help
in understanding what these quantum effects might be.
The other parameter is $n$, describing the fall-off of the
dark energy as a function in $r$. In \cite{hess2009} $n=2$ was
assumed, but this already violates observations in the solar system. From that, calculations were performed with
$n=3$ \cite{book}, however, in \cite{nielsen1,nielsen2}
the inspiral wave of a gravitational wave was fitted to 
pcGR and concluded that $n$ has to be larger than 3.
From then on, the value $n=4$ is assumed, which is also in 
better accordance to the $r$-dependence obtained in 
\cite{visser}. Nevertheless, one
has to keep in mind that $B_n$ and $n$ are parameters, 
converting pcGR to a phenomenological theory. There exist
studied for the accumulation of {\it dark matter}
in a black hole. dark matter does not interact strongly with 
matter, only through gravitation. In contrast,
dark energy does interact with matter \cite{visser} and
should play a more dominant role.

The metric in pcGR, including rotation, has the
form \cite{PPNP} (there is an error in the
$g_{11}$-component, corrected here)

\beqa
g_{00} & = & -\frac{ r^2 - 2m_0 r  + a^2 \cos^2 \vartheta + \frac{B_n}{(n-1)(n-2)r^{n-2}} }{r^2 + a^2 \cos^2\vartheta}~~~, \nonumber \\
g_{11} & = & \frac{r^2 + a^2 \cos^2 \vartheta}{r^2 - 2m_0 r + a^2 +  \frac{B_n}{(n-1)(n-2)r^{n-2}}  }~~~, \nonumber \\
g_{22} & = &  r^2 + a^2 \cos^2 \vartheta~~~,  \nonumber \\
g_{33} & = & (r^2 +a^2 )\sin^2 \vartheta + \frac{a^2 \sin^4\vartheta \left(2m_0 r -  \frac{B_n}{(n-1)(n-2)r^{n-2}}  \right)}{r^2 + a^2 \cos^2 \vartheta}~~~,  \nonumber \\
g_{03} & = & \frac{-a \sin^2 \vartheta ~ 2m_0 r + a \frac{B_n}{(n-1)(n-2)r^{n-2}}   \sin^2 \vartheta }{r^2 + a^2 \cos^2\vartheta}
~~~.
\label{eq8}
\eeqa

It is illustrative to calculate the energy-momentum tensor
within the pc-Schwarzschild limit (non-rotating star).
Substituting
the metric (\ref{eq8}) into the left hand side of the Einstein 
equations
results in the $T_{\mu\nu}$ on the right hand side of these
equations. Further,
for the energy-momentum tensor we define the dimensionless
quantities $y=\frac{r}{m_0}$
and $b=\frac{B}{m_0^4}$, with which this tensor acquires 
the form (the signature of the metric is (-+++))

\beqa
8\pi \left(T^\mu_{~\nu} \right) & = & 
\left(
\begin{array}{cccc}
-\frac{b}{2y^6} & 0 & 0 & 0 \\
0 & -\frac{b}{2y^6} & 0 & 0 \\
0 & 0 & \frac{b}{y^6} & 0 \\
0 & 0 & 0 & \frac{b}{y^6} \\
\end{array}
\right)
~=~ 8 \pi
\left(
\begin{array}{cccc}
-\varrho^\Lambda & 0 & 0 & 0 \\
0 & p^\Lambda_r & 0 & 0 \\
0 & 0 & p^\Lambda_\vartheta & 0 \\
0 & 0 & 0 & p^\Lambda_\vartheta \\
\end{array}
\right)
~~~.
\label{eq6c}
\eeqa
From (\ref{eq6c}) the Riemann scalar $R=-\frac{b}{y^6}$
is deduced.
It is clearly seen that the energy-momentum tensor
describes an anisotropic fluid with a different pressure
in the radial compared to the angular directions.

From these consideration, it should be clear that the
pseudo-complex extension of GR has to be considered seriously.
The next question is: What kind of prediction the pcGR
is making and in what they differ to GR? 

\section{Resum\'e of some predictions of pcGR}

\begin{figure}[ht]
\begin{center}
\rotatebox{270}{\resizebox{150pt}{200pt}{\includegraphics[width=0.23\textwidth]{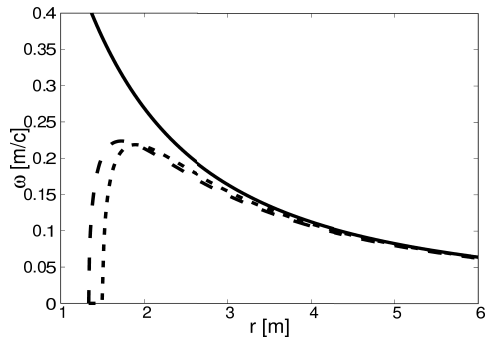}}}
\end{center}
\caption{ 
The orbital frequency of a point particle in a circular orbit 
as a function in $r$. The rotational parameter
is $a=0.95m_0$. The upper curve is the result as
obtained in GR while the two lower curves are from pc-GR,
for $n=3$ (left curve) and $n=4$ (right curve).
}
 \label{fig1}
\end{figure}

\begin{center}
\begin{figure}[ht]
\begin{center}
\rotatebox{0}{\resizebox{400pt}{400pt}{\includegraphics[width=0.23\textwidth]{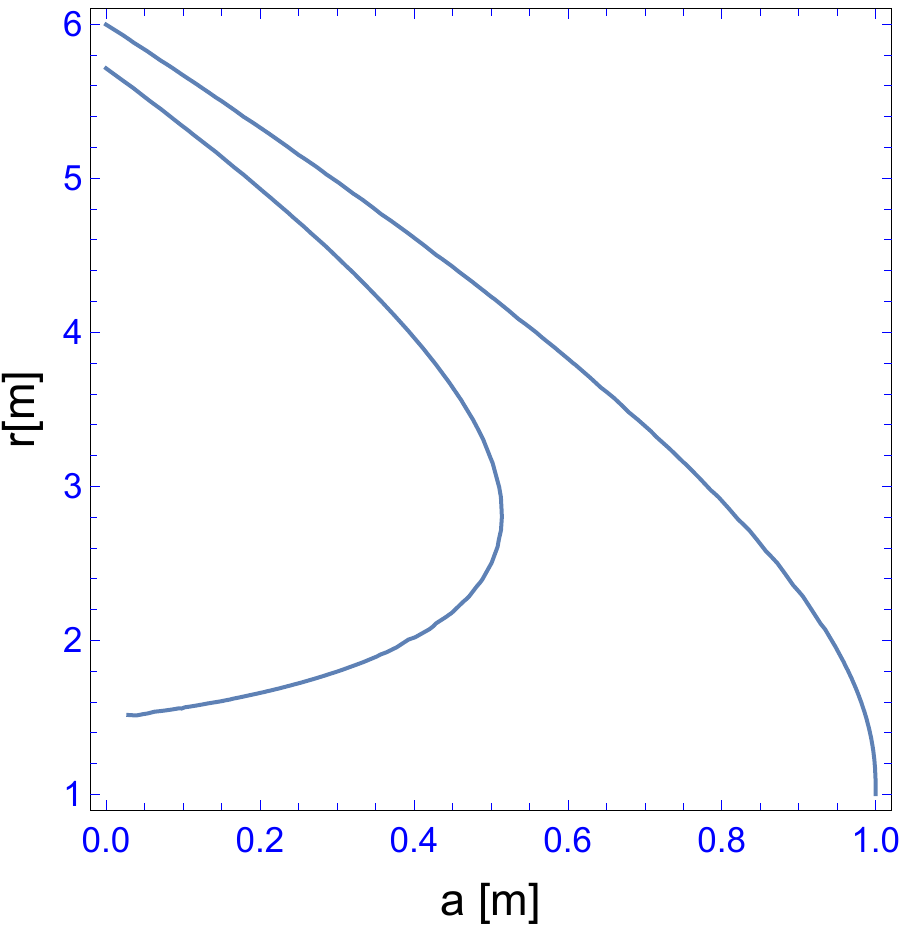}}}
\end{center}
\caption{ 
Limits for stable orbits ($n=4$) as a function in the
rotational parameter $a$. 
The upper curve is the
result of GR, which marks the position of the
{\it Innermost Stable Circular Orbit} (ISCO). In pc-GR
there are no stable orbits to the left of the lower curve.
}
 \label{fig2}
\end{figure}
\end{center}

\begin{figure}[ht]
\begin{center}
\rotatebox{0}{\resizebox{200pt}{200pt}{\includegraphics[width=0.23\textwidth]{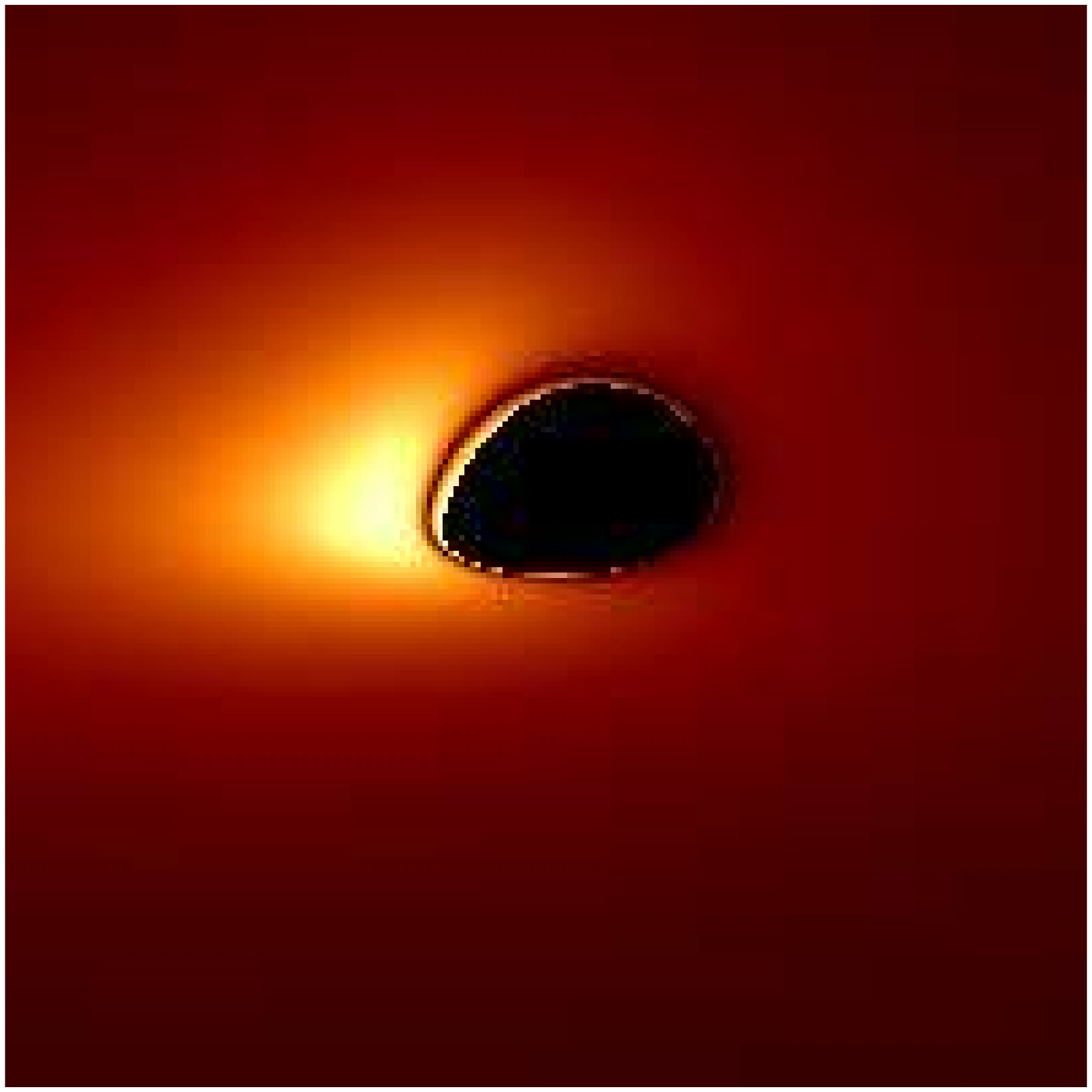}}}
\rotatebox{0}{\resizebox{200pt}{200pt}{\includegraphics[width=0.23\textwidth]{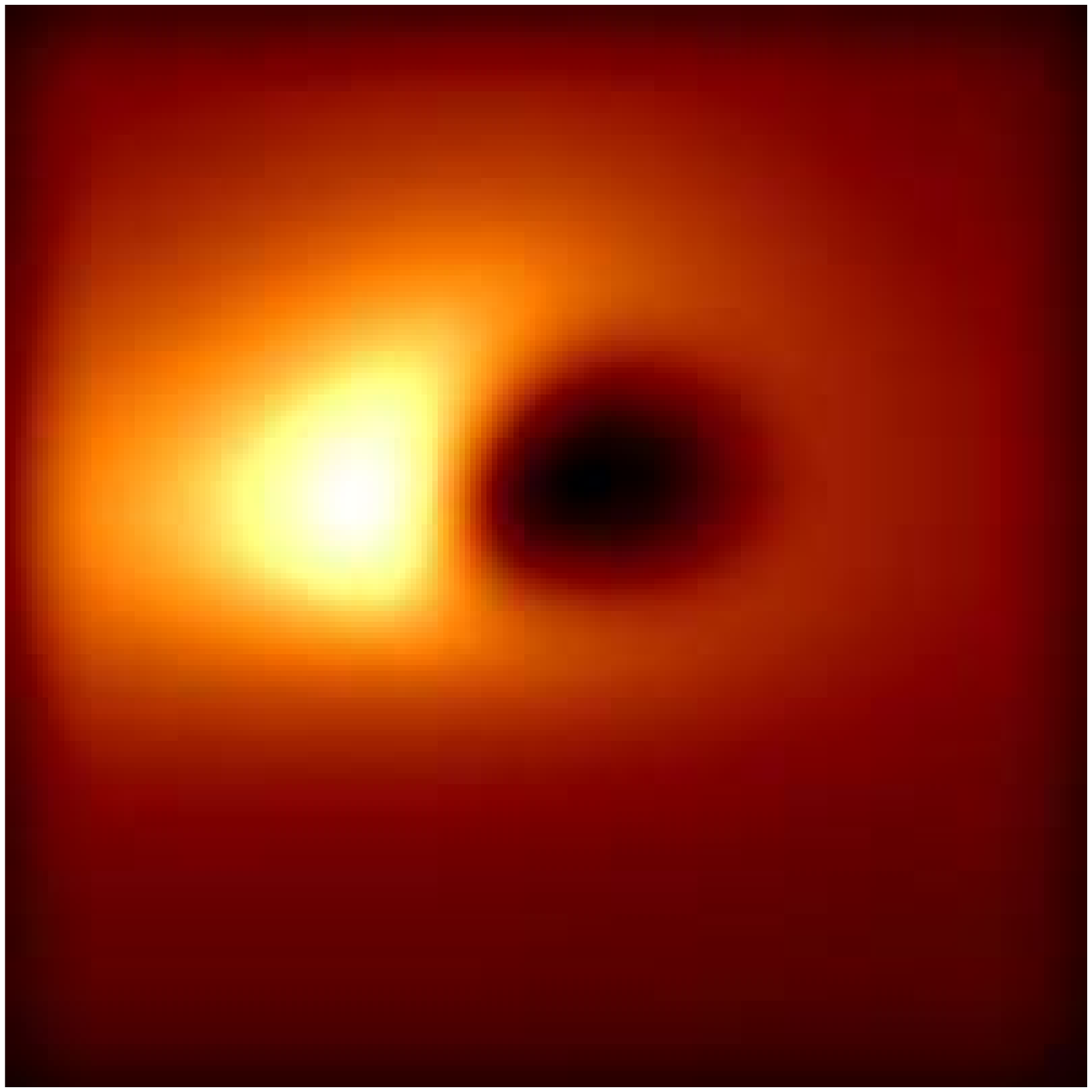}}}
\end{center}
\caption{ 
Simulations of the accretion disc in M87, using the model of
Ref. \cite{page} and $a=0.95m_0$. The left panel
is for a resolution of 5$\mu$as. The right panel
is for 20$\mu$ as, a little bit less than
24$\mu$as, the resolution reported by the EHT.
The inclination angle of the disc with respect to the observer
is 70$^o$.
}
\label{fig3}
\end{figure}

\subsection{Structure of accretion disks}

In \cite{book,universe,MNRAS1,MNRAS2} simulations of
a thin, optical thick accretion disk \cite{page} 
where published and compared to GR. 

In order to understand the results of the simulations, 
we have to
explain how the orbital frequency of a particle in 
a circular orbit changes with respect to the radial
distance and where do stable orbits exist: In Fig. \ref{fig1}
this orbital frequency is depicted for the Kerr parameter
$a=0.95m_0$. Clearly seen is the appearance of a maximum,
at which two neighboring orbitals have the same orbital 
frequency, which results in less friction and less emission
if light, marking the position of the dark ring. 
In Fig. \ref{fig2} the range of stable orbits
is depicted \cite{MNRAS2014}. The upper line corresponds 
to GR, where for $a=0$ the last stable orbit is at $6m_0$
and it approaches $1m_0$ for $a=1m_0$. The lower curve 
encircles the area where no stable orbit exists in pcGR. 
Note that above $a=0.5m_0$ all orbits are stable in pcGR
and they can reach the point where the orbital frequency has
a maximum. Below $a=0.5m_0$ the pcGR follows the GR	1,
i.e., similar results are expected, though, the last stable
orbit is further in in more light is emitted.

In Fig. \ref{fig3}, 
on the left panel the simulation for a high resolution 
is given. It is clearly observed that the disk has a dark ring,
followed further in by a bright one. The dark ring is the
consequence of the maximum in the orbital frequency as 
a function on the radial distance. The right panel
in Fig. \ref{fig3} shows the same but with a low resolution
of 20$\mu$as, a bit lower than the EHT. One can see that
the ring structure is washed out. Thus, in order to see
the ring structure one has to increase the resolution
of the EHT significantly. It also means that the
observations are consistent with GR {\it and} pcGR. 
Thus, this particular
prediction of pcGR cannot be verified yet and one has to
await a much better resolution.

\subsection{Redshift}

\begin{figure}[ht]
\begin{center}
\rotatebox{0}{\resizebox{250pt}{160pt}{\includegraphics[width=0.23\textwidth]{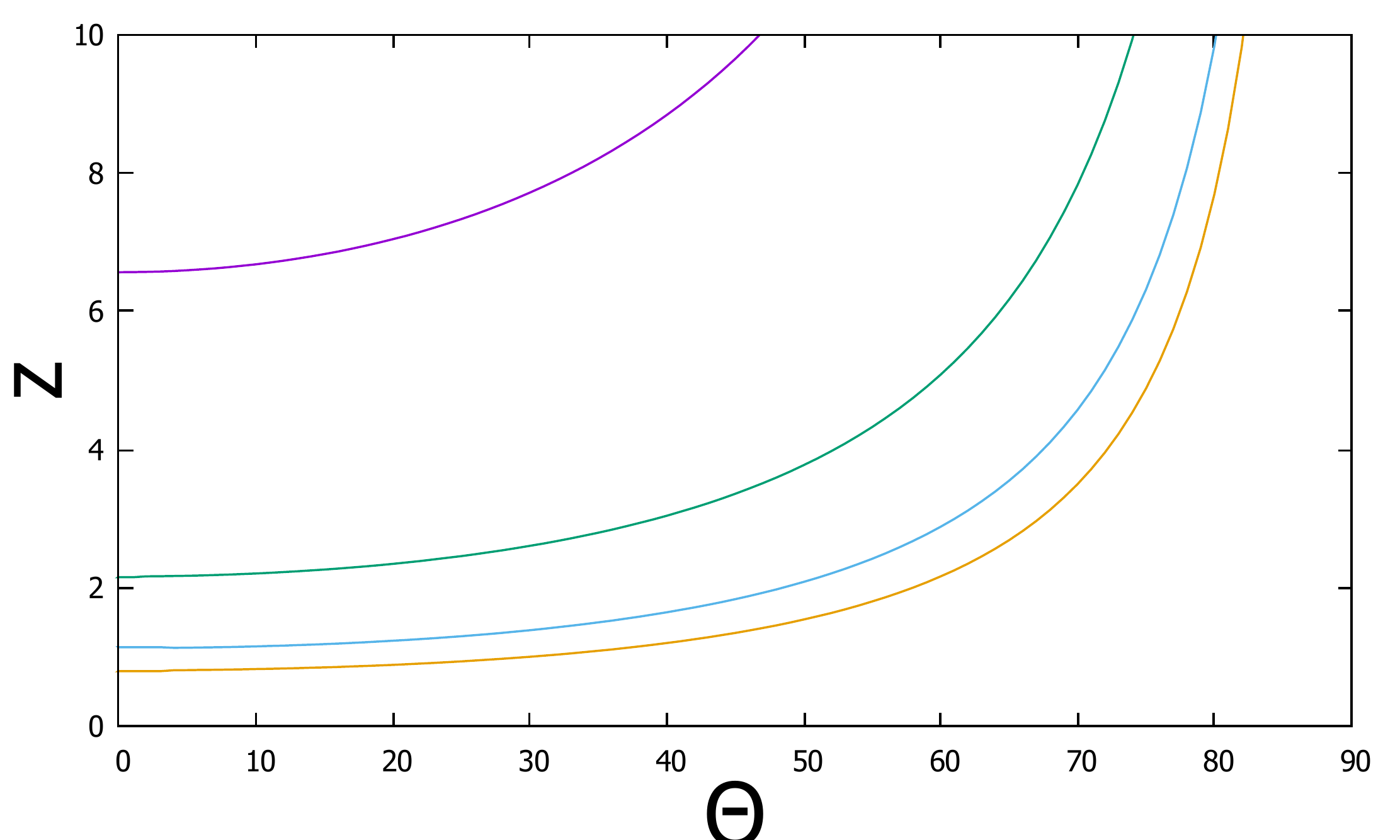}}}
\end{center}
\caption{ 
Redshift $z$ at $r=\frac{3}{2}m_0$ for different rotation
parameter values $a$, as a function on the azimuth
angle $\theta$. From the upper
to the subsequent lower curves
the $a$-value varies from 0.2$m_0$, 0.5$m_0$, 0.8$m_0$ 
to 1.0$m_0$.
}
 \label{fig4}
\end{figure}

There is hopefully a possibility to see some difference between
GR and pcGR, but it depends on the non-existence of a
jet. Because the jet is a consequence of the presence of
an accretion disk, this is equivalent to demand that there
is no accretion disk. 

The only observable black hole up to now by the EHT, which may have no accretion disk, is SgrA* in the center of our Galaxy.
Observational results are still pending.

In Fig. \ref{fig4} the redshift {\it at the surface of the
black star} is plotted as a function on the
azimuthal angle $\theta$. The different curves correspond to
different $a$-values. The redshift is lowered toward the
poles and if $a$ is very large (above 0.5$m_0$), the
redhift has values of 1-2. In other words, infalling
matter, which collides with the surface, should emit light
and its reshifted frequency has to 
be detected by an observer on earth. 
Toward the orbital plane the redshift is too large and no hope 
exist to detect it. I.e., there is only hope to see something{
near the poles, which requires that, first, 
the matter has to fall
onto the surface near the poles and, secondly, there is
no jet present (due to an accretion disk), which would
over-shine the effect.

\subsection{Neutron Stars}

In the first decade of this century a neutron star with  2.05 solar masses was observed \cite{neutron-2.05}, while until then
one did not expect masses larger than 2,
with a "realistic" equations of state. In 
\cite{dexheimer} a theory for neutron stars was 
developed which finally could accommodate the mass. Recently,
in observing gravitational waves \cite{neutron-2.5}, 
a further candidate with 2.5 solar masses was suggested. Since then, many contributions can be found which can explain these
masses. The central point is how to change the equation of state
of the matter in order to support a greater mass.
This seems a bit unsatisfactory, because greater masses can
obviously obtained by changing appropriately 
the equation of state, such that larger masses can be obtained.

\begin{figure}[ht]
\begin{center}
\rotatebox{0}{\resizebox{230pt}{200pt}{\includegraphics[width=0.5\textwidth]{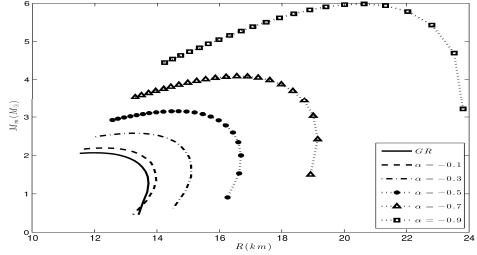}}}
\end{center}
\caption{Mass vs the radius of a black star. A linear relation
between the dark energy density and mass density in the interior
of the star was assumed.
}
\label{fig5}
\end{figure}

In pcGR this is not a problem: The accumulation 
of dark energy, due to the coupling of the mass to it,
accommodates well larger masses, as is shown in 
\cite{book,isaac}. There, the coupling of the mass density with
the dark energy density was assumed to be linear

\beqa
\varrho_\Lambda & = & \alpha \varrho_{\rm m} 
~~~.
\label{eq9}
\eeqa
The index ${\rm m}$ refers to the mass density and the index
$\Lambda$ to the dark energy density. Also here the mass
couples to the dark energy. In fig. \ref{fig5} several
curves for different $\alpha$-values are depicted. Stars
up to 6 solar masses were found stable. However, if these
dark stars can still be considered neutron stars is questionable.
In our view, there is a continuous shift from a
classical neutron star, with a particle beam, to a classical
black hole, where the emitted light of a beam 
shifts to a large redshift or disappears
due to unknown processes.

In \cite{caspar} semi-classical calculations \cite{birrell}
were performed which resulted in a more complicated 
relation for the coupling of the matter density to the
dark energy density. As a result, the coupling diminished
toward to the surface of the black star and much higher
masses were obtained.

All calculations relied on the hadronic model for the matter,
published in \cite{dexheimer}. With increasing density this
model reaches its limit and due to lack of confidence
no further calculations can
be done. In this respect, new models for the matter at high
density are required. 

Therefore, we
predict the existence of stellar objects which behave as
a neutron star, emitting a beam whose emitted light has
a large redshift, and also
have too large masses to be explained by standard models.  

\subsection{Quasi Periodic Objects (QPO)}

\begin{figure}[ht]
\begin{center}
\rotatebox{0}{\resizebox{200pt}{200pt}{\includegraphics[width=0.5\textwidth]{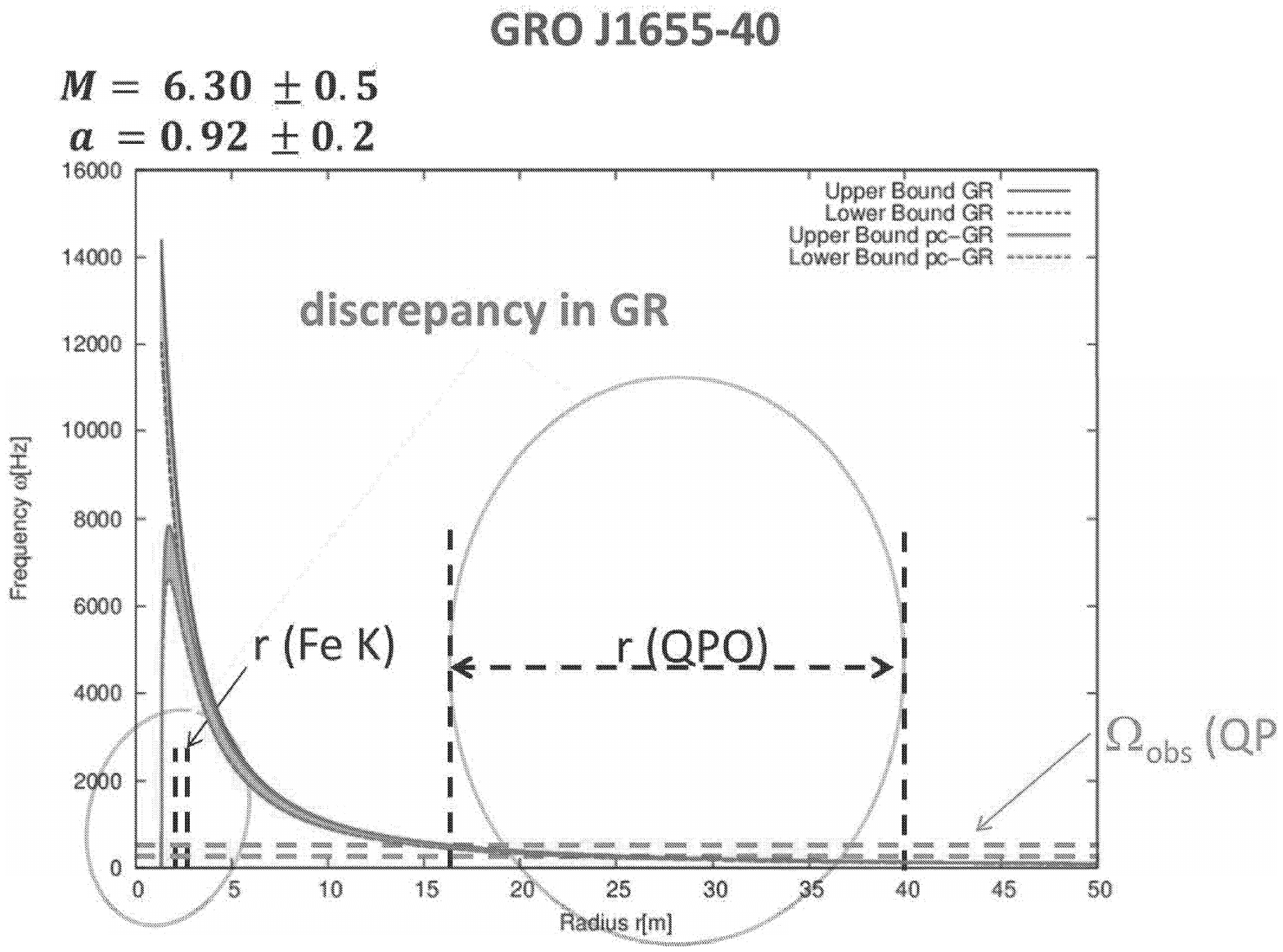}}}
\rotatebox{0}{\resizebox{200pt}{200pt}{\includegraphics[width=0.5\textwidth]{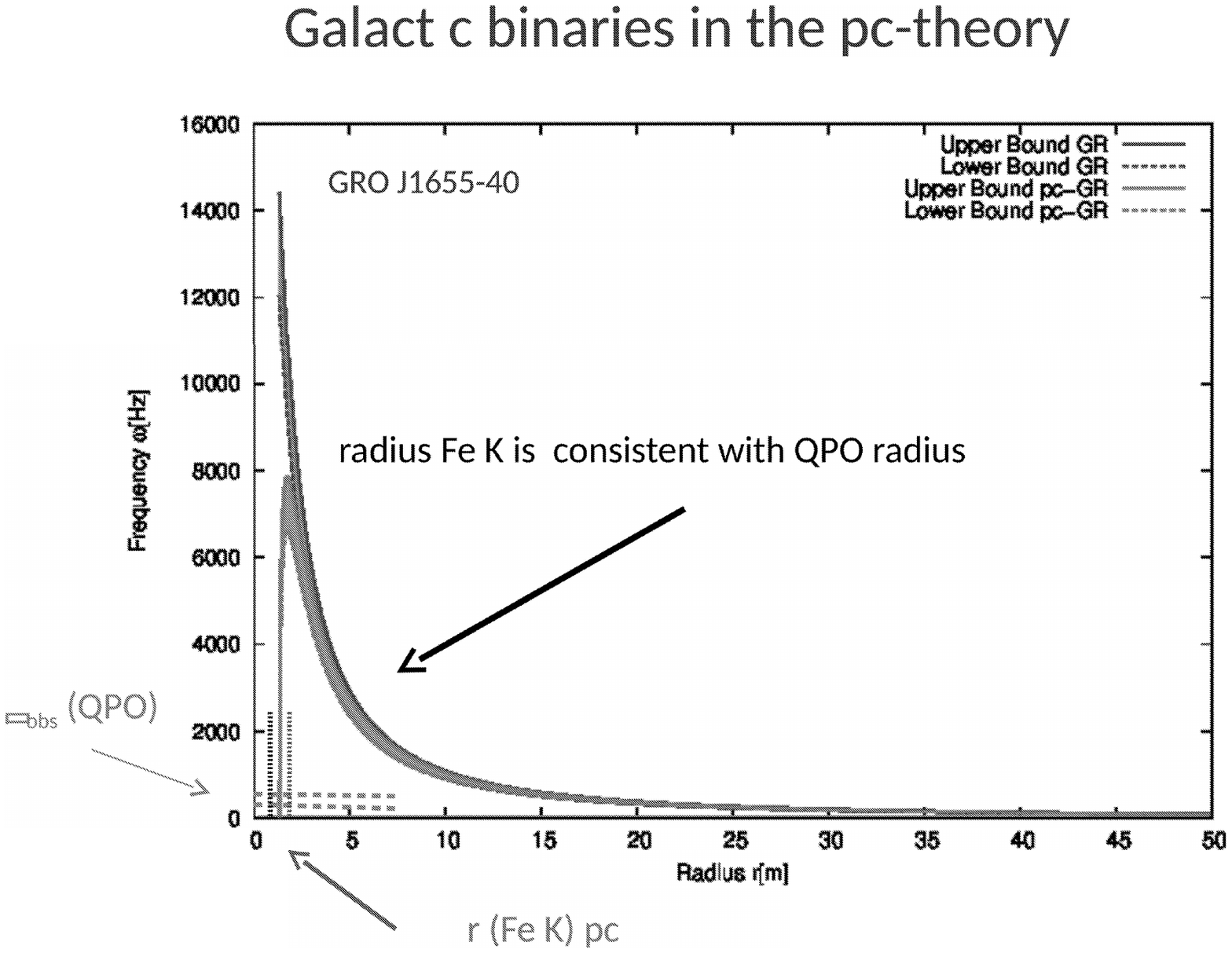}}}
\end{center}
\caption{The orbital frequency of a particle in a circular orbit
versus the radial distance. The upper curve is for GR while
the lower one for pcGR and $n=3$ (for $n=4$ the result is
not much different). The width of a line is a measure of
the error in the $a$-parameter. The horizontal dashed
line is from fitting to the orbital frequency observed, with
its experimental error. The vertical dashed line is for
adjusting to the redshift of the emitted line. 
}
\label{fig6}
\end{figure}

QPO's are assumed to be 
excitations within an accretion disks (hot spots)
which are circling around with the disk, They are observed
in accretion disks of galactic black holes and also
in the disks of stellar black holes
\cite{QPO1,QPO2,QPO3,QPO4}. Knowing the frequency
and the formula for the orbital frequency in a circular orbit,
a distance value can be deduced. If this is the real distance
has to be verified by another measurement, as the redshift
of the line emitted. Unfortunately, for galactic black holes
this has not been seen yet. Because galactic black holes
are fairly well isolated (no large partner nearby),
the interpretation of the QPO as a co-rotating 
hot spot in the accretion disk is quite well established.
For stellar black holes one also observes QPO's and 
simultaneously the
rotation frequency and the redshift of the emitted line.
Therefore, from the redshift one can deduce also a distance.
{\it From both observations the same distance has to result}, 
for consistency. 
In Fig. \ref{fig6} the orbital frequency of a particle in
a circular orbit is depicted. The upper line refers to GR while
the lower line to pcGR. The horizontal dashed line
is from adjusting to the observed orbital frequency and
the vertical line from the observed redshift of the emitted 
line.

The problem is now, 
using the formulas from GR both distances
are {\it not the same}, but they {\it are the same} in pcGR.
One might say that this proves pcGR, however, one has to take
into account possible effects of the stellar partner 
onto the accretion disk around the 
black hole.
It was shown that one can indeed explain the discrepancy
in GR by this influence of the stellar partner, using models
\cite{QPO-model}.
Though pcGR is easier to apply here, not involving further
assumptions, the fact that one can reconcile the measurement
with GR makes it inconclusive.

\section*{Conclusions}

I have presented the motivation for the algebraic extension
of GR to pseudo-complex coordinates. The pcGR turns out
to be the {\it only viable algebraic extension}.

The structure of pcGR was explained and its differences
to GR pointed out.

A theory has to be verifiable, which is exactly the
case for pcGR. Several predictions were presented: 

\begin{itemize}

\item The appearance of a dark ring followed further in by 
a bright ring, which is due to the dependence of the
orbital frequency of a particle in a circular orbit. 
Unfortunately, the resolution of the EHT is not sufficient
to distinguish pcGR from GR, until a much better resolution
is reached.

\item The redshift of light emitted from the surface of
the dark star, as a function of the azimuthal angle
was determined. 
We predict that matter falling onto the surface near the
poles can emit light with a redshift between 1-2, which
should be observable. However, this is linked to the 
condition that the massive object is not accompanied
by an accretion disk, which would include a jet near the
poles. This jet would over-shine the effect described. 
The black hole in the center of our galaxy, SgrA*, 
is thought not to have an accretion disk, so there is
some hope.

\item The theory allows a stable star for any mass 
value, i.e., it easily accommodates larger masses for
neutron stars, without any further assumptions,
using an equation of state known.
Possibly, there is a smooth transition from stellar
object which behave as a neutron star with a beam,
to stellar objects where the beamdisappears.
The uncertainty here is the structure
(equation of state) within the star. 

\item Quasi Periodic Objects, which are observed in
galactic centers and in binary systems with one black
hole. Here, the simultaneous measurement of 
the orbital frequency and the redshift of the emitted line
lead to {\it the same distance}, contrary to GR.
However, investigating the effect of the stellar visible
partner on the structure of the accretion disk, using models,
can still accommodate GR. This is a case where pcGR
leads to a simple explanation, but still the situation is not
settled yet.

\end{itemize}

As can be seen, the pcGR is a realistic extension of GR,
but further observations with their improvements have to
be made.

\section*{Acknowledgment}
Financial support from DGAPA-PAPIIT (IN100421} is 
acknowledged.

\end{document}